\documentclass[aps,prl,reprint,footinbib]{revtex4-1} 
\usepackage{bm, braket,amsmath,mathtools,comment,color}
\usepackage{graphicx}
\usepackage{hyperref}
\begin{document}

\title{Quasiexact Kondo Dynamics of Fermionic Alkaline-Earth-Like Atoms at Finite Temperatures}

\author{Shimpei Goto}
\email[]{goto@phys.kindai.ac.jp}
\author{Ippei Danshita}
\email[]{danshita@phys.kindai.ac.jp}
\affiliation{Department of Physics, Kindai University, Higashi-Osaka city, Osaka 577--8502, Japan}

\date{\today}

\begin{abstract}
A recent experiment has observed the antiferromagnetic interaction between the ground state \({}^1S_0\) and the metastable state \({}^3P_0\) of \(^{171}\)Yb atoms, which are fermionic. This observation combined with the use of state-dependent optical lattices allows for quantum simulation of the Kondo model. We propose that in this Kondo simulator the anomalous temperature dependence of transport, namely the Kondo effect, can be detected through quench dynamics triggered by the shift of a trap potential. For this purpose, we improve the numerical efficiency of the minimally entangled typical thermal states (METTS) algorithm by applying additional Trotter gates. Using the improved METTS algorithm, we compute the quench dynamics of the one-dimensional Kondo model at finite temperatures quasi-exactly. We find that the center-of-mass motion exhibits a logarithmic suppression with decreasing the temperature, which is a characteristic feature of the Kondo effect.
\end{abstract}
\maketitle

Orbital degrees of freedom are a fundamental element for understanding physics of various condensed matter systems, including heavy-fermion materials~\cite{steglich_superconductivity_1979,stewart_heavy-fermion_1984}, transition metal oxides~\cite{falicov_simple_1969,kugel_polaron_1980,tokura_orbital_2000}, iron pnictides~\cite{kamihara_iron-based_2008,kuroki_unconventional_2008,si_high-temperature_2016}, and compound semiconductors~\cite{byrnes_excitonpolariton_2014,kunes_excitonic_2015}. In these systems, the multiorbital character, together with the spin degrees of freedom and strong interparticle interactions, leads to the emergence of magnetism, superconductivity, excitonic condensation, and the Kondo effect. It is widely believed that the essence of some of these properties can be extracted by analyzing the two-orbital Anderson- and Kondo-type models, in which delocalized fermions in one orbital exchanges their spins with localized fermions in the other orbital. However, accurate simulation of these models with classical resources is in general intractable because of the exponential growth of the Hilbert space and the minus sign problem in quantum Monte Carlo simulations.

An alternative approach for analyzing the prototypical two-orbital models is analog quantum simulation~\cite{feynman_simulating_1982} using optical lattices loaded with ultracold gases~\cite{bloch_quantum_2012,gross_quantum_2017,hofstetter_quantum_2018}. 
It has been proposed that optical-lattice quantum simulators (OLQSs) of the two-orbital models can be realized with use of fermionic alkaline-earth-like atoms (AEAs)~\cite{gorshkov_two-orbital_2010,foss-feig_probing_2010, nakagawa_laser-induced_2015, zhang_kondo_2016, cheng_enhancing_2017}, such as strontium~\cite{desalvo_degenerate_2010} and ytterbium~\cite{fukuhara_degenerate_2007, taie_realization_2010}. 
A remarkable advantage of AEAs over alkali atoms is the existence of the electronically excited state \({}^3P_0\) or \({}^3P_2\) with long lifetime, which can be coupled to the ground state \({}^1S_0\) via an ultranarrow clock transition. Riegger {\it et al.}~indeed have used a state-dependent optical lattice to create a two-orbital fermionic quantum gas of \(^{173}\)Yb~\cite{riegger_localized_2018}, in which atoms in \({}^1S_0\) (\({}^3P_0\)) state play a role of delocalized (localized) fermions. 
Moreover, Ono {\it et al.}~have reported the observation of antiferromagnetic spin-exchange interaction between the \({}^1S_0\) and \({}^3P_0\) states of \(^{171}\)Yb~\cite{ono_antiferromagnetic_2019}. Since \(^{171}\)Yb atoms in the \({}^1S_0\) state hardly interact with each other~\cite{kitagawa_two-color_2008}, their two-orbital system in a state-dependent optical lattice naturally simulates the Kondo model.

One of the most important targets of the OLQS of the Kondo model is the Kondo effect~\cite{kondo_resistance_1964,abrikosov_electron_1965,yosida_bound_1966,anderson_poor_1970,wilson_renormalization_1975,andrei_solution_1983}, in which a localized fermion forms a many-body spin-singlet state with delocalized fermions when the temperature is lowered. 
The formation of such Kondo singlets causes the anomalous increase of the resistance with decreasing the temperature. The Kondo effect is thought to be a key concept for understanding rich quantum phases and phase transitions of the Kondo lattice model represented by a Doniach phase diagram~\cite{doniach_kondo_1977}. 
Since transport properties of trapped quantum gases have been often investigated by measuring their center-of-mass (COM) motion induced in response to a sudden displacement of the trapping potential~\cite{burger_superfluid_2001, fertig_strongly_2005, strohmaier_interaction-controlled_2007, mckay_phase-slip-induced_2008, haller_pinning_2010, tanzi_velocity-dependent_2016,boeris_mott_2016}, it is likely that the Kondo effect in the OLQS of the Kondo model will be detected via such simple transport measurements. 
However, accurate theoretical predictions on the COM dynamics have never been made because of the difficulties in calculating real-time evolution of the quantum many-body system with two orbitals at finite temperatures.

In this letter, we develop a numerical method that overcomes such difficulties in order to show that the Kondo effect of the Kondo OLQS can be indeed detected by measuring the COM motion of the delocalized fermions after the trap displacement. Specifically, we restrict ourselves to one-dimensional (1D) systems, in which matrix product states (MPS) serve as an efficient description of states with relatively low energy~\cite{white_density_1992,white_density-matrix_1993,schollwock_density-matrix_2011}, and modify the finite-temperature algorithm using the minimally entangled typical thermal states (METTS)~\cite{white_minimally_2009,stoudenmire_minimally_2010}. 
Our modified METTS algorithm includes additional Trotter gates and allows for efficient simulations of systems with an Abelian symmetry, such as the Hubbard and Kondo models. 
Using the modified METTS, we compute the finite temperature dynamics of the Kondo model with the antiferromagnetic interaction and find that when the temperature decreases, the maximum COM speed during the dynamics logarithmically decreases, i.e., the transport exhibits a logarithmic suppression, which is a characteristic feature of the Kondo effect. 
We also analyze the fully spin-polarized system and the ferromagnetic Kondo model, in which the Kondo effect does not occur~\cite{furusaki_kondo_2005}, as references to be compared with the antiferromagnetic case. 
The logarithmic suppression of the transport is found to be absent in these two cases.

\textit{Model and method.---}
We consider an ultracold mixture of \(^{171}\)Yb atoms, which are fermionic, in the \({}^1S_0\) and \({}^3P_0\) states confined in a combined potential of optical lattices and a parabolic trap. 
We assume that the transverse optical lattice is sufficiently deep such that the system can be regarded to be spatially 1D. 
The longitudinal optical lattice is state dependent in a way that an atom in the \({}^3P_0\) state is localized at \(j=0\) while the lattice for \({}^1S_0\) atoms is modestly deep for the tight-binding approximation to be valid but is shallow enough to make \({}^1S_0\) atoms delocalized. 
This system is well described by the 1D Kondo model~\cite{kasuya_theory_1956,kondo_resistance_1964} with a parabolic trap term,
\begin{equation}
  \label{eq:hamiltoinan}
  \begin{aligned}
  \hat{H} = &-J\sum^{L-1}_{i = -L} \sum_\sigma\left(\hat{c}^\dagger_{i \sigma} \hat{c}_{i+1 \sigma} + \mathrm{H.c.} \right) + V \hat{\bm{s}}_{i=0}\cdot \hat{\bm{S}}_\mathrm{imp}\\ 
  &+ w \sum^{L}_{i=-L}\sum_\sigma {(i-x_\mathrm{c}/a)}^2\hat{n}_{i\sigma},
  \end{aligned}
\end{equation}
and can be regarded as an OLQS of the model.
The total number of sites is \(2L+1\). Here, \(\hat{c}^\dagger_{i\sigma}\) (\(\hat{c}_{i\sigma}\)) creates (annihilates) a \(^1S_0\) fermion with spin \(\sigma \) at site \(i\), and \(\hat{n}_{i \sigma} = \hat{c}^\dagger_{i\sigma}\hat{c}_{i \sigma}\). 
\(\hat{\bm{s}}_i = (\hat{s}^x_i, \hat{s}^y_i, \hat{s}^z_i) \) are spin operators of a \(^1S_0\) fermion at site \(i\) and each component is defined as \(\hat{s}^\gamma = (1/2)\sum_{\alpha \beta} \hat{c}^\dagger_{i \alpha}\sigma^\gamma_{\alpha \beta}\hat{c}_{i\beta}\) with the Pauli matrices \(\bm{\sigma}^\gamma \).
\(\hat{\bm{S}}_\mathrm{imp} = (\hat{S}^x_\mathrm{imp}, \hat{S}^y_\mathrm{imp}, \hat{S}^z_\mathrm{imp})\) are spin operators of the impurity fermion at site \(0\).
\(J \) denotes the hopping amplitude of \(^1S_0\) fermions,  
\(V\) the spin-exchange interaction between \(^1S_0\) and \(^3P_0\) fermions, \(w\) the amplitude of the trap, \(x_\mathrm{c}\) the position of the trap center, and \(a\) the lattice constant. 

\begin{figure}
  \includegraphics[width=\linewidth]{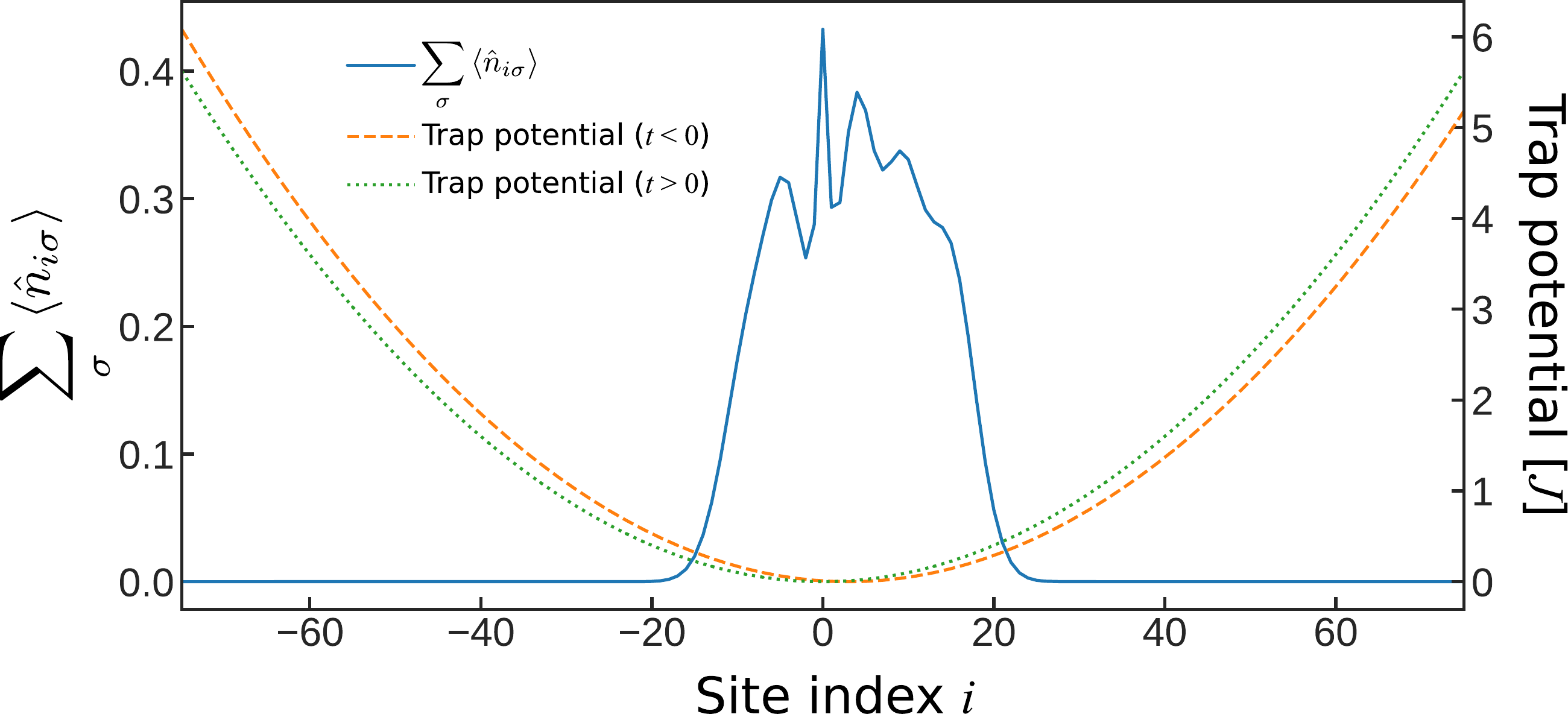}
  \caption{(Color online) The solid line represents the density distribution of the delocalized fermions of the Kondo model in the stationary state for \(N=9\), total magnetization \(M=0\), \(V/J = 1\), \(x_c = 3a\), and \(\beta J = 75\). The localized fermion is located at \(i = 0\). The dashed and dotted lines represent the parabolic trap potential before and after the displacement of its center.\label{fig:schematic}}
\end{figure}

The interaction between \(^{171}\)Yb atoms in the \(^1S_0\) state can be safely ignored because it is very small (the \(s\)-wave scattering length is \(a_s=-0.15\,{\rm nm}\)~\cite{kitagawa_two-color_2008}).
It is worth noting that there exists direct interaction between \(^1S_0\) and \(^3P_0\) fermions, which is given by \(V_\mathrm{d} (\sum_\sigma \hat{n}_{i=0\sigma})\hat{n}_\mathrm{imp}\)~\cite{ono_antiferromagnetic_2019}. 
Since the number of a \(^3P_0\) fermion \(\hat{n}_\mathrm{imp}\) is fixed to be unity, the direct interaction is equivalent to a barrier potential at site \(0\).
We assume that a laser beam is focused on site 0 to cancel the direct interaction.
Such control can be made in experiment, e.g., by using a digital micromirror device~\cite{parsons_site-resolved_2016,mazurenko_cold-atom_2017}.
With this Hamiltonian~\eqref{eq:hamiltoinan}, we calculate the time evolution of the COM position 
\(\hat{x}_\mathrm{G} = \sum^L_{i=-L,\sigma} i a \hat{n}_{i\sigma}/N\)
with total particle number of \(^1S_0\) fermions \(N = \sum_{i\sigma}\braket{\hat{n}_{i\sigma}}\) and the COM velocity 
\(\hat{v}_\mathrm{G} = -\frac{\mathrm{i}}{\hbar}[\hat{x}_\mathrm{G}, \hat{H}]\)
followed by the shift of a trap center \(x_\mathrm{c}\) from \(3a\) to \(0\) at finite temperatures as depicted in Fig.~\ref{fig:schematic}.

In order to numerically simulate dynamics of quantum many-body systems at finite temperatures, we use MPS and the METTS algorithm. There is another option for computing such finite-temperature dynamics, namely, the purification method~\cite{verstraete_matrix_2004,feiguin_finite-temperature_2005,binder_minimally_2015,goto_cooling_2017}. However, since in the purification method the density matrix of a system is represented as a pure state by squaring the dimensions of local Hilbert spaces, it is not very efficient for our two-orbital system with large local Hilbert spaces.

In the METTS algorithm, thermal expectation value at an inverse temperature \(\beta = 1/ k_\mathrm{B}T \) is calculated as 
\begin{equation}
  \label{eq:thermal_expectation}
  \begin{aligned}
    \braket{\hat{O}}_\mathrm{\beta} &= \frac{\mathrm{Tr}\left[\mathrm{e}^{-\frac{\beta}{2}\hat{H}} \hat{O} \mathrm{e}^{-\frac{\beta}{2}\hat{H}}\right]}{Z} \\
    &=  \sum_i \frac{\Braket{i|\mathrm{e}^{-\beta \hat{H}}|i}}{Z}\frac{\Braket{i|\mathrm{e}^{-\frac{\beta}{2}\hat{H}} \hat{O} \mathrm{e}^{-\frac{\beta}{2}\hat{H}}|i}}{\Braket{i|\mathrm{e}^{-\beta \hat{H}}|i}},
  \end{aligned}
\end{equation}
and the summation over orthonormal basis \(\ket{i}\) is performed by the Markov-chain Monte Carlo (MCMC) sampling.
In the ordinary METTS algorithm~\cite{white_minimally_2009,stoudenmire_minimally_2010}, the transition probability of MCMC method from a state \(\ket{i}\) to \(\ket{j}\) is given by
\begin{align}
  \label{eq:prob_metts}
  p_{i \to j} = \frac{\left|\Braket{j|\mathrm{e}^{-\frac{\beta}{2}\hat{H}}|i}\right|^2}{\Braket{i|\mathrm{e}^{-\beta \hat{H}}|i}}
\end{align}
With this transition probability, the METTS algorithm suffers from a severe autocorrelation problem at high temperatures as easily inferred from the \(\beta \to 0\) limit.
This autocorrelation problem can be eliminated by breaking the total particle number conservation.
However, the breaking of the conservation of particle number leads to significant increase of computation time in dynamics.
Hence, we introduce the following transition probabilities for odd steps,
\begin{equation}
  \label{eq:prob_odd}
  p^\mathrm{odd}_{i \to j} = \frac{\left|\Braket{j|{\left[\hat{U}(\tau)\right]}^n\mathrm{e}^{-\frac{\beta}{2}\hat{H}}|i}\right|^2}{\Braket{i|\mathrm{e}^{-\beta \hat{H}}|i}}
\end{equation}
and 
\begin{equation}
  \label{eq:prob_even}
  p^\mathrm{even}_{i \to j} = \frac{\left|\Braket{j|\mathrm{e}^{-\frac{\beta}{2}\hat{H}}{\left[\hat{U}^\dagger(\tau)\right]}^n|i}\right|^2}{\Braket{i|{\left[\hat{U}(\tau)\right]}^n\mathrm{e}^{-\beta \hat{H}}{\left[\hat{U}^\dagger(\tau)\right]}^n|i}}
\end{equation}
for even steps.
Here, \(\tau \) is a parameter which characterizes the Trotter gates,
\begin{equation}
  \label{eq:Trotter}
  \hat{U}(\tau) = \mathrm{e}^{-\mathrm{i}\tau\hat{H}_\mathrm{even}}\mathrm{e}^{-\mathrm{i}\tau\hat{H}_\mathrm{odd}},
\end{equation}
and \(n\) is an integer.
For \(\hat{H}_\mathrm{even}\) and \(\hat{H}_\mathrm{odd}\), one can take any hermitian operators as long as they respect the conservation of particle number.
We use the products of local Hilbert states in some symmetric sector as the orthonormal basis \(\ket{i}\). This approach is a variant of the symmetric METTS algorithm~\cite{binder_symmetric_2017} with symmetric bases \(\ket{i}\) and \({\left[ \hat{U}^\dagger(\tau)\right]}^n \ket{i}\), which are very flexible because of the parameter \(\tau \) and the freedom to choose \(\hat{H}_\mathrm{even}\) and \(\hat{H}_\mathrm{odd}\).
Moreover, the implementation of our approach is quite easy since it requires only the applications of the Trotter gates in addition to the ordinary METTS algorithm.
With the transition probabilities, we can reduce the autocorrelation time by increasing \(\tau \) and \(n\) without breaking the conservation.
However, since the bond dimensions of MPS increase with \(\tau \) and \(n\), some tuning of the parameters may be required for efficient simulations.
The validity of our approach and some benchmark results are shown in Supplemental Material~\footnote{See Supplemental Material at [URL will be inserted by publisher] for the validity and some benchmark results of our improved method, the comparison of the Kondo temperatures we define and the ordinary perturbative one, and the derivation of the relation between the resistance \(R\) and the quantity \(\tilde{R}\)}.

The dynamics of thermal expectation value is obtained by representing the operator \(\hat{O}\) in the Heisenberg picture \(\hat{O}(t) = \mathrm{e}^{\mathrm{i} t \hat{H}^\prime/\hbar} \hat{O} \mathrm{e}^{-\mathrm{i}t \hat{H}^\prime / \hbar}\) with the Hamiltonian \(\hat{H}^\prime \) after a quench.
Both imaginary and real time evolutions of MPS in this study are performed with the time-evolving block decimation method~\cite{vidal_efficient_2003,vidal_efficient_2004,white_real-time_2004,daley_time-dependent_2004} using the optimized Forest-Ruth-like decomposition~\cite{omelyan_optimized_2002}.
Throughout this study, we set \( w/J = 0.001 \), the number of delocalized fermions \(N\) to nine, \(L=75\), \(\tau = 1.0/J\), \(n=4\), and \(\hat{H}_\mathrm{even}\) (\(\hat{H}_\mathrm{odd}\)) is the Hamiltonian linking even (odd) bonds of the Kondo model~\eqref{eq:hamiltoinan}. On-site terms are equally divided to \(\hat{H}_\mathrm{even}\) and \(\hat{H}_\mathrm{odd}\).
Truncation error is set to \(10^{-10}\) in imaginary-time evolution and \(10^{-8}\) in real-time evolution, and the bond dimensions are allowed to increase up to \(4000\).

\begin{figure}
  \includegraphics[width=\linewidth]{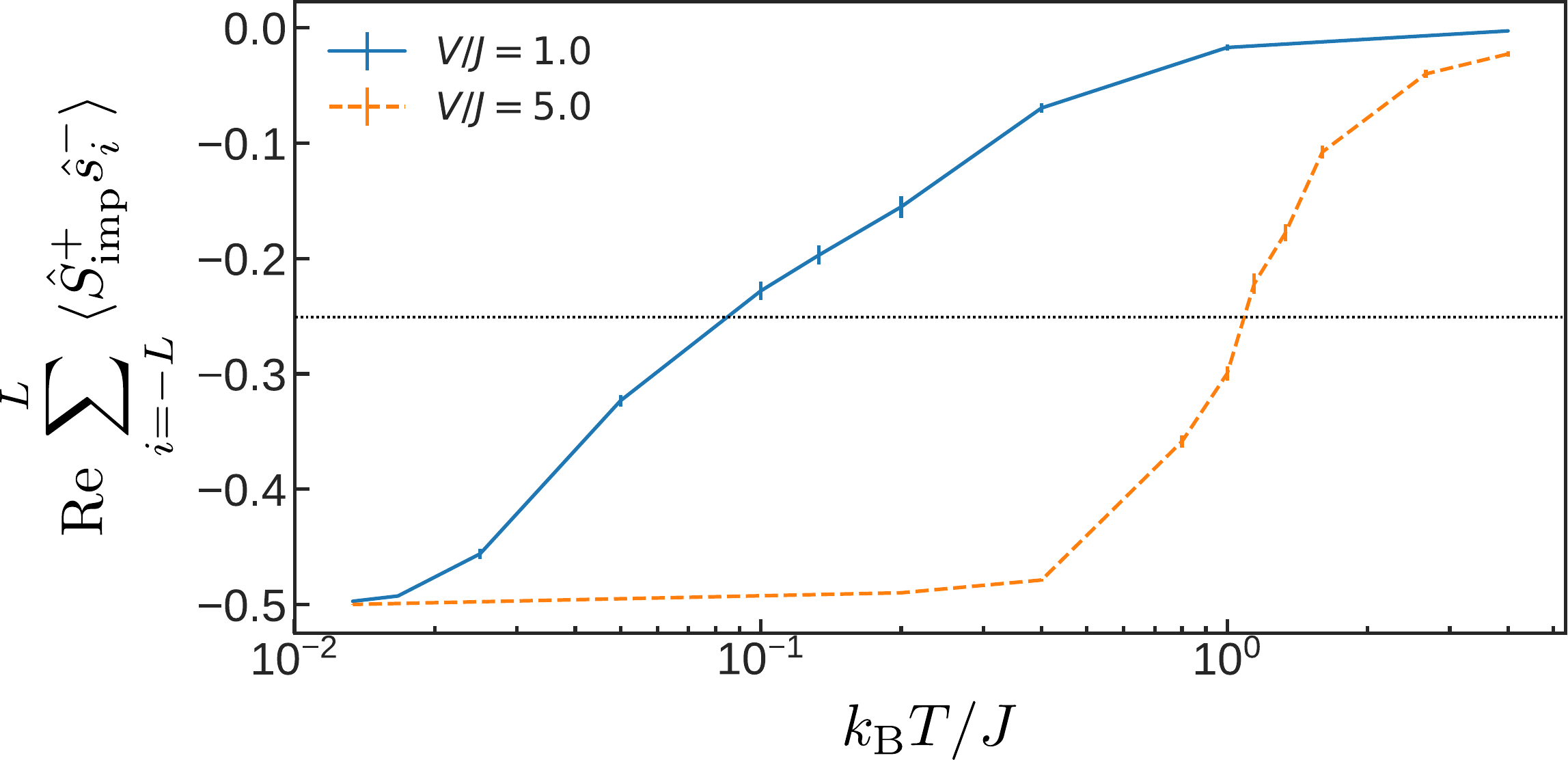}
  \caption{(Color online) The spin correlation versus temperature in steady states for \(V/J = 1\) (blue solid line) and \(5\) (orange dashed line).
  Error bars indicate 1\(\sigma \) uncertainty.
  The horizontal dotted line represents \(\mathrm{Re}\sum_i\braket{\hat{S}^+_\mathrm{imp}\hat{s}^-_i} = -0.25\), the criterion we used to define the Kondo temperature.
  We set \(M=0\) and \(x_\mathrm{c} = 0\).\label{fig:s_corr}}
\end{figure}

\textit{Antiferromagnetic case.---} We first consider the case that the spin exchange interaction is antiferromagnetic and the total magnetization is zero, i.e., \(V>0\) and \(M = \Braket{\hat{S}^z_\mathrm{imp} + \sum_i \hat{s}^z_i} = 0\). 
In order to identify a temperature range in which the Kondo effect occurs, we show in Fig.~\ref{fig:s_corr} the spin correlation,  \(\mathrm{Re}\sum_i\braket{\hat{S}^+_\mathrm{imp}\hat{s}^-_i} = \sum_{i}\braket{ \hat{S}^x_\mathrm{imp}\hat{s}^x_i + \hat{S}^y_\mathrm{imp}\hat{s}^y_i}\), for \(V/J = 1\) (blue solid line) and \(5\) (orange dashed line), as a function of the temperature. It is clearly seen that when the temperature decreases, the spin correlation grows logarithmically, implying the formation of a many-body spin-singlet.

From these spin correlations, we can extract an important energy scale called Kondo temperature \(T_\mathrm{K}\). At \(T \lesssim T_\mathrm{K}\) the spin-singlet correlation grows such that it makes significant contributions to physical quantities.
In this study, we define the Kondo temperature as a temperature at the spin correlation \(\mathrm{Re} \sum_i \braket{\hat{S}^+_\mathrm{imp}\hat{s}_i}\) becomes \(-0.25\), namely the half of the maximal singlet value.
As discussed in Supplemental Material~\cite{Note1}, the Kondo temperatures given by this definition behave similarly to the ordinary Kondo temperatures obtained by the perturbative renormalization group analysis~\cite{hewson_kondo_1997} at least around \(V/J = 1\).
The Kondo temperature at \(V/J=1\) is around \(0.1 J / k_\mathrm{B}\) and the estimated size of the Kondo screening cloud is around \(10 a\)~\cite{Note1,sorensen_scaling_1996}.
In the following calculations for real-time dynamics at \(V/J = 1\), we take \(7.5 \leq \beta J \leq 75.0\), which corresponds approximately to \(0.1 \leq T/ T_\mathrm{K} \leq 1\).
Moreover, itinerant atoms are distributed over 40 sites (See Fig.~\ref{fig:schematic}), which is sufficiently larger than the Kondo screening length. 
Thus, our setting is adequate for observing the Kondo physics.

Notice that the Kondo temperature \(T_\mathrm{K} \sim 0.1 J/k_\mathrm{B}\) at \(V/J\) is remarkably lower than the lowest temperature, \(T = 0.25J/k_{\rm B}\), achieved in experiments with ultracold fermions~\cite{mazurenko_cold-atom_2017}. 
We emphasize that the Kondo temperature can be significantly lifted by increasing \(V/J\).
For instance, Fig.~\ref{fig:s_corr} shows that \(T_\mathrm{K}\) at \(V/J = 5\) is well above \(T = 0.25J/k_{\rm B}\).
Nevertheless, we set \(V/J = 1\) for computations of real-time dynamics because the numerical cost is much more expensive for higher temperatures.

\begin{figure}
  \includegraphics[width=\linewidth]{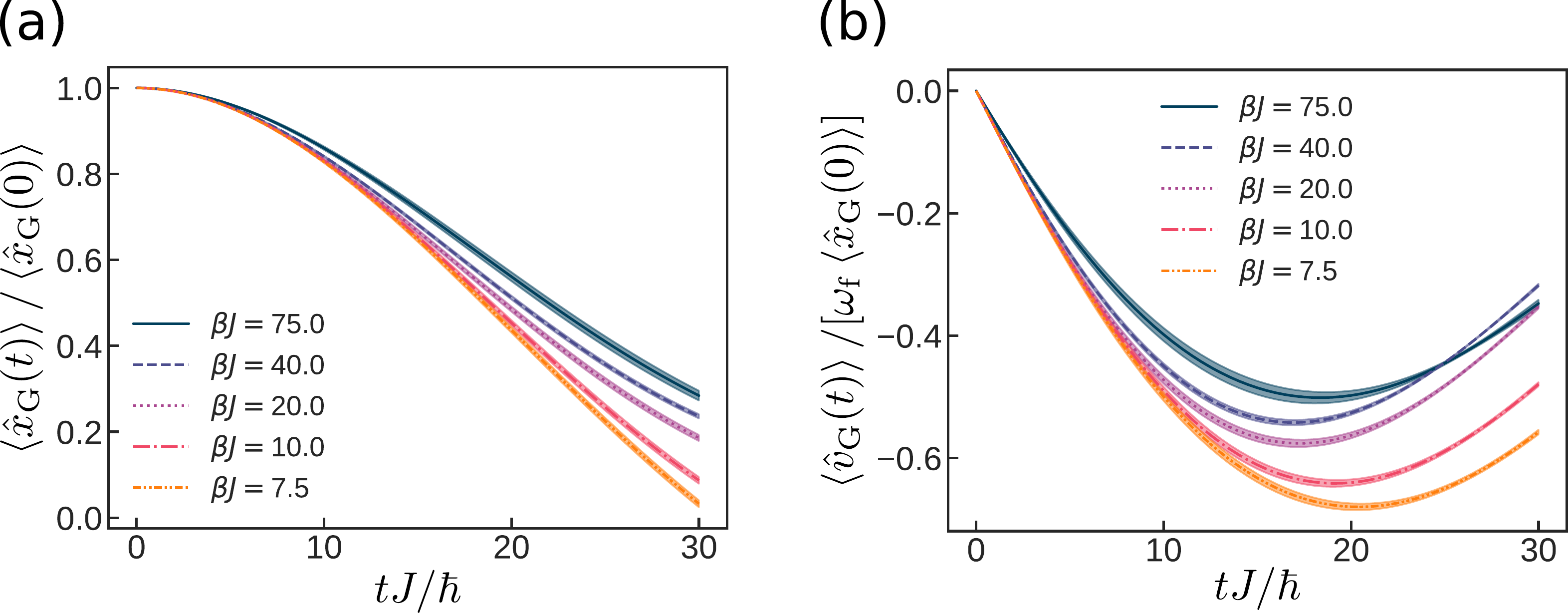}
  \caption{(Color online) Time evolution of the normalized COM positions (a) and velocities (b) for several temperatures. We set \(V/J = 1\) and \(M=0\). The shaded regions are 1\(\sigma \) uncertainty of trajectories.\label{fig:AntiFerro}}
\end{figure}

Figure~\ref{fig:AntiFerro} shows the time evolution of the COM positions and velocities at several temperatures after the shift of the trap center from \(x_\mathrm{c} = 3a\) to \(x_\mathrm{c} = 0\).
The COM positions and velocities are respectively normalized by \(\braket{\hat{x}_{\rm G}(0)}\) and \(\omega_{\rm f} \braket{\hat{x}_{\rm G}(0)}\), where \(\omega_{\rm f}=2\sqrt{wJ}/\hbar \) denotes the dipole oscillation frequency of free particles~\cite{rey_ultracold_2004}. \(\omega_{\rm f} \braket{\hat{x}_{\rm G}(0)}\) means the maximum speed during the undamped dipole oscillation starting with the position \(\braket{\hat{x}_{\rm G}(0)}\).
In Fig.~\ref{fig:AntiFerro}, we see that the transport is significantly suppressed when the temperature decreases. This tendency is reminiscent of the Kondo effect, in which the resistance increases with decreasing the temperature.

\begin{figure}
  \includegraphics[width=\linewidth]{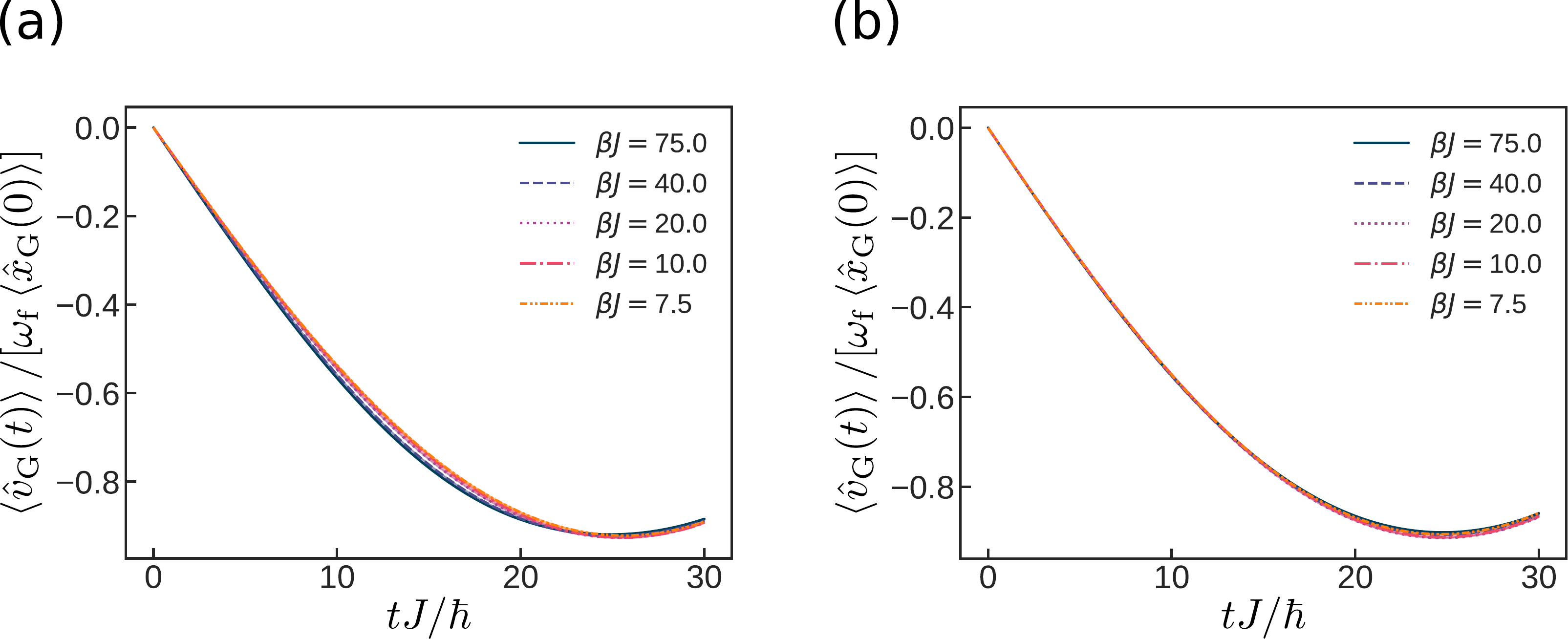}
  \caption{(Color online) Time evolution of the normalized COM velocities in the fully spin-polarized system with \(V/J = 1\)  (a) and in the ferromagnetic Kondo model with \(V/J = -1\) (b).~\label{fig:pol_and_ferro}}
\end{figure}

\textit{Fully spin-polarized and ferromagnetic cases.---} In order to support our argument that the suppression of transport is a manifestation of the Kondo effect, we also compute the quench dynamics of the following two systems, which are widely known NOT to exhibit the Kondo effect~\cite{furusaki_kondo_2005}.
The first example is a fully spin-polarized system (\(M=5\)), in which the spin-exchange interaction term in Eq.~\eqref{eq:hamiltoinan} acts as a simple potential barrier term.
In this system, we completely prohibit spin-flip processes that are essential for the Kondo effect~\cite{kondo_resistance_1964}.
Figure~\ref{fig:pol_and_ferro} (a) represents the dynamics of the normalized COM velocities in the fully spin-polarized system.
Except the total magnetization \(M\), any other settings are equivalent to those of the dynamics shown in Fig.~\ref{fig:AntiFerro}.
In contrast to the \(M=0\) case in Fig.~\ref{fig:AntiFerro} (b), the normalized velocities in Fig.~\ref{fig:pol_and_ferro} (a) does not show visible temperature dependence.
This behavior is consistent with the formula of the resistance, \(R\propto T^{2K-2}\), obtained from the Tomonaga-Luttinger (TL) liquid theory, where \(K\) denotes the Luttinger parameter for the charge sector and \(K=1\) for noninteracting fermions~\cite{kane_transmission_1992, kagan_supercurrent_2000, buchler_superfluidity_2001, danshita_universal_2013}.

The second example is the case that the spin-exchange interaction is ferromagnetic. Specifically, we take \(V/J=-1\) and \(M=0\).
Notice that while there exists the ferromagnetic Kondo effect in 1D for \(K < 1\)~\cite{furusaki_kondo_1994,furusaki_kondo_2005}, this is not the case for non-interacting delocalized fermions considered here. This happens because they are also described by the TL liquid theory with \(K=1\). 
Figure~\ref{fig:pol_and_ferro} (b) shows the time dependence of the normalized COM velocities in the ferromagnetic Kondo model.
Except the sign of \(V/J\), any other settings are equivalent to the settings in Fig.~\ref{fig:AntiFerro}.
Likewise the fully spin-polarized case, the normalized COM velocities in the ferromagnetic Kondo model do not exhibit visible temperature dependence and this is also consistent with \(R\propto T^{2K-2}\).

\begin{figure}
  \includegraphics[width=\linewidth]{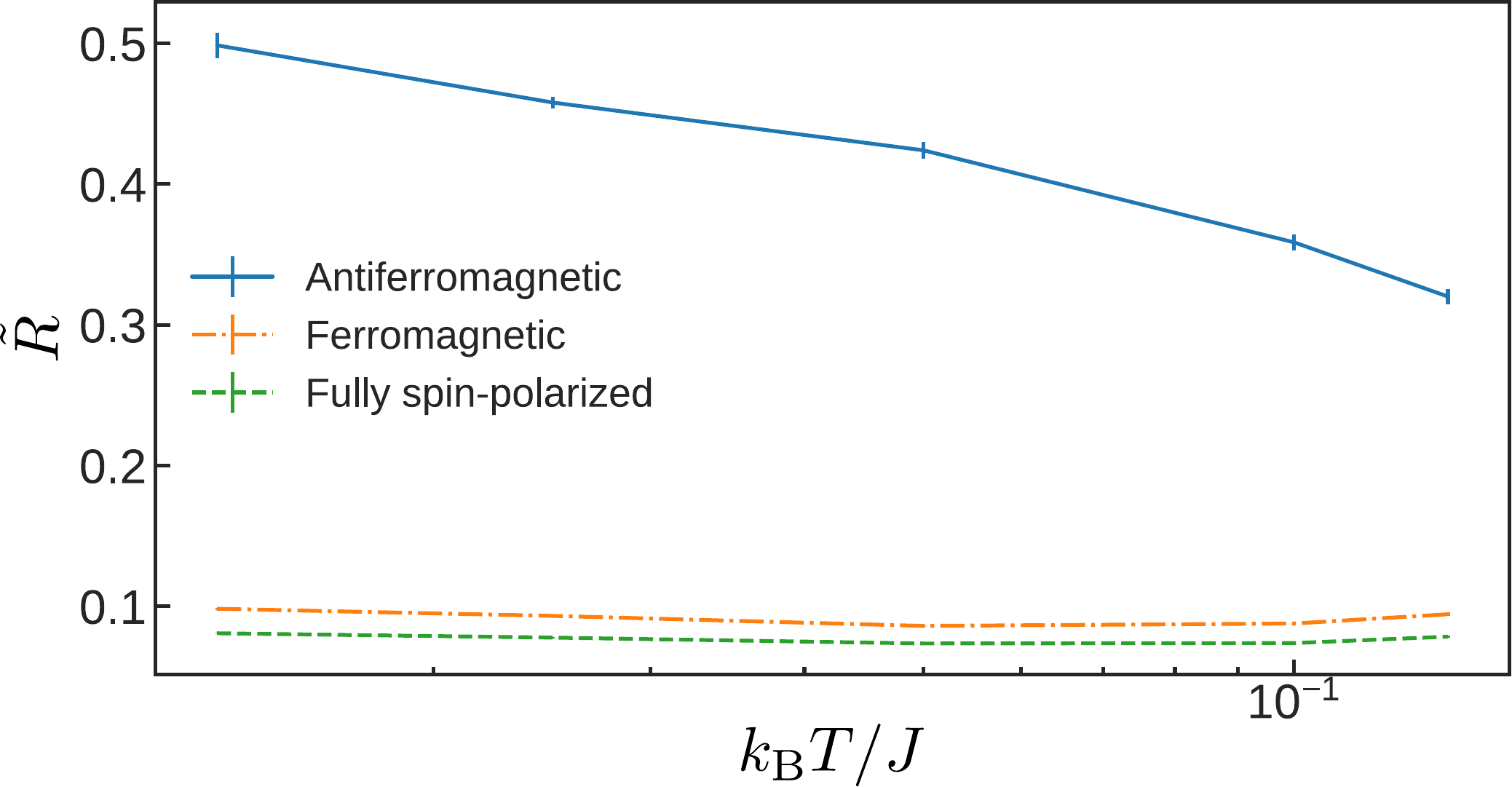}
  \caption{(Color online) Temperature dependence of \(\tilde{R}= 1-\max_t |\langle\hat{v}_{\rm G}(t) \rangle |/(\omega_{\rm f}\langle\hat{x}_{\rm G}(0) \rangle)\). The solid, dashed, dashed-dotted lines represent the antiferromagnetic Kondo  (\(M=0\) and \(V/J = 1\)), fully spin-polarized (\(M=5\) and \(V/J = 1\)), and ferromagnetic Kondo (\(M=0\) and \(V/J = -1\)) systems. Error bars indicate 1\(\sigma \) uncertainty.~\label{fig:peak}}
\end{figure}

In order to discuss the temperature dependence of the transport more quantitatively, we plot the quantity, \(\tilde{R} = 1-\max_t |\braket{\hat{v}_{\rm G}(t)}|/[\omega_{\rm f}\braket{\hat{x}_{\rm G}(0)}]\), in Fig.~\ref{fig:peak}. 
As shown in the Supplemental Material~\cite{Note1}, \(\tilde{R}\) is approximately proportional to the resistance \(R\) under the assumption that \(\tilde{R}\ll 1\), and is suited to characterizing the transport.
In Fig.~\ref{fig:peak}, we see that the temperature dependence of the transport is only visible in the antiferromagnetic Kondo model with spin-flip processes (blue solid line).
We emphasize that the horizontal axis of Fig.~\ref{fig:peak} is logarithmic scale; \(\tilde{R}\) of the antiferromagnetic Kondo model exhibits a logarithmic growth with decreasing the temperature, which is a characteristic feature of the Kondo effect.
Specifically, the ``resistance'' \(\tilde{R}\) increases by around 1.7 times when the temperature decreases from \( T \sim T_\mathrm{K}\) to \(0.1 T_\mathrm{K}\).
Strictly speaking, we observe only the lower temperature side \(0.1 \lesssim T/T_\mathrm{K} \lesssim 1\) of the expected logarithmic dependence~\cite{costi_kondo_2000,merker_conductance_2013}.
For the higher temperature side \(1 \lesssim T/T_\mathrm{K} \lesssim 10\), it requires very expensive numerical cost.
In this sense, corroborating the logarithmic dependence in the higher temperature region will be a suitable target of OLQS experiments.
We suggest that observation of the logarithmic temperature dependence serves as a smoking-gun signature of the Kondo effect in the OLQS of the Kondo model. 

\textit{Summary.---}
In order to propose an experimental way for observing the Kondo effect with ultracold alkaline-earth-like atoms in optical lattices, we numerically simulated the finite temperature dynamics of the one-dimensional Kondo model with using quasi-exact minimally entangled typical thermal states (METTS) algorithm based on matrix product states.
We found that when the spin-exchange interaction is antiferromagnetic, the COM motion after a sudden displacement of the trap potential is suppressed logarithmically with decreasing the temperature. 
In contrast, it was shown that such suppression of the transport is absent in the ferromagnetic Kondo model or the fully spin-polarized system.
These findings convincingly indicate that the Kondo effects in ultracold atoms are detectable via the simple transport measurement.

We also improved the numerical efficiency of the METTS algorithm without breaking the total particle number conservation by the applications of the Trotter gates. 
The modified METTS algorithm can be applied to other systems for accurately analyzing static and dynamical properties at finite temperatures, such as the spectral functions~\cite{coira_finite-temperature_2018} and the out-of-time ordered correlations~\cite{bohrdt_scrambling_2017}.

\begin{acknowledgments}
We thank K. Ono and Y. Takahashi for fruitful discussions.
The MPS calculations in this work are performed with ITensor library, http://itensor.org.
This work was financially supported by KAKENHI from Japan Society for the Promotion of Science: Grant No. 18K03492 and No. 18H05228, by  CREST, JST No. JPMJCR1673, and by MEXT Q-LEAP Grant No.\@ JPMXS0118069021.
\end{acknowledgments}

\end{document}


\title{Supplemental Material for ``Quasiexact Kondo Dynamics of Fermionic Alkaline-Earth-Like Atoms at Finite Temperatures''}
\author{Shimpei Goto}
\email[]{goto@phys.kindai.ac.jp}
\author{Ippei Danshita}
\email[]{danshita@phys.kindai.ac.jp}
\affiliation{Department of Physics, Kindai University, Higashi-Osaka city, Osaka, Japan}
\date{\today}
\maketitle
\section{Validity of transition probabilities with Unitary transformations}
In the main text, we introduce the transition probabilities for odd Markov steps
\begin{align}
  \label{eq:prob_odd}
  p^\mathrm{odd}_{i \to j} = \frac{\left|\Braket{j|\hat{U}\mathrm{e}^{-\frac{\beta}{2}\hat{H}}|i}\right|^2}{\Braket{i|\mathrm{e}^{-\beta \hat{H}}|i}}
\end{align}
and 
\begin{align}
  \label{eq:prob_even}
  p^\mathrm{even}_{i \to j} = \frac{\left|\Braket{j|\mathrm{e}^{-\frac{\beta}{2}\hat{H}}\hat{U}^\dagger|i}\right|^2}{\Braket{i|\hat{U}\mathrm{e}^{-\beta \hat{H}}\hat{U}^\dagger|i}}
\end{align}
for even steps.
Here, a state \(\ket{i}\) is some orthonormal basis, \(\beta \) is inverse temperature, \(\hat{H}\) is a system Hamiltonian, and \(\hat{U}\) is some unitary operator.
The stationary distribution of the Markov chain should be the canonical ensemble
\begin{align}
    \label{eq:canonical}
\Pi_i = \frac{\Braket{i|\mathrm{e}^{-\beta \hat{H}}|i}}{Z}
\end{align}
so that the minimally entangled typical thermal states (METTS) algorithm gives correct thermal expectation values~\cite{white_minimally_2009,stoudenmire_minimally_2010}.
Here, \(Z = \sum_i\braket{i|\mathrm{e}^{-\beta \hat{H}}|i}\).
In this section of the Supplemental Material, we show the stationary distribution of the Markov chain defined by the transition probabilities~\eqref{eq:prob_odd} and~\eqref{eq:prob_even} is the canonical ensemble~\eqref{eq:canonical}.

The stationary distribution can be obtained as the left eigenvector with the eigenvalue 1 of a transition matrix defined as \(p_{i,j} = p_{i\to j}\)~\cite{gubernatis_quantum_2016}.
By taking the two Markov steps as one step, the transition probability for every two step can be given as 
\begin{align}
    p_{i \to j} = \sum_k p^\mathrm{odd}_{i \to k} p^\mathrm{even}_{k \to j}.
\end{align}
With this transition probability, one can directly confirm that the canonical ensemble \(\Pi_i\)~\eqref{eq:canonical} is the stationary distribution as follows:
\begin{align}
    \begin{aligned}
    \sum_i\Pi_i p_{i \to j} &= \frac{1}{Z}\sum_{i, k} \left|\Braket{k|\hat{U}\mathrm{e}^{-\frac{\beta}{2}\hat{H}}|i}\right|^2 \frac{\left|\Braket{j|\mathrm{e}^{-\frac{\beta}{2}\hat{H}}\hat{U}^\dagger|k}\right|^2}{\Braket{k|\hat{U}\mathrm{e}^{-\beta \hat{H}}\hat{U}^\dagger|k}} \\
    &= \frac{1}{Z}\sum_k \Braket{k|\hat{U}\mathrm{e}^{-\beta \hat{H}}\hat{U}^\dagger|k}\frac{\left|\Braket{j|\mathrm{e}^{-\frac{\beta}{2}\hat{H}}\hat{U}^\dagger|k}\right|^2}{\Braket{k|\hat{U}\mathrm{e}^{-\beta \hat{H}}\hat{U}^\dagger|k}} \\
    &= \frac{1}{Z}\sum_k \Braket{j|\mathrm{e}^{-\frac{\beta}{2}\hat{H}}\hat{U}^\dagger|k}\Braket{k|\hat{U}\mathrm{e}^{-\frac{\beta}{2}\hat{H}}|j} \\
    & = \Pi_j.
    \end{aligned}
\end{align}
Therefore, the canonical ensemble \(\Pi_i\)~\eqref{eq:canonical} is the left eigenvector with the eigenvalue 1 and thus the stationary distribution of the Markov chain defined by the transition probabilities~\eqref{eq:prob_odd} and~\eqref{eq:prob_even}.

\section{A benchmark result in the Kondo model}
In this section, we show that the unitary transformation implemented by the successive applications of the Trotter gates used in the main text reduces the autocorrelation of samples and the total computation time through a performance test with the Kondo model,
\begin{align}
  \label{eq:hamiltoinan}
  \hat{H} = -J\sum^{L-1}_{i = -L} \sum_\sigma\left(\hat{c}^\dagger_{i \sigma} \hat{c}_{i+1 \sigma} + \mathrm{H.c.} \right) + V \hat{\bm{s}}_{i=0}\cdot \hat{\bm{S}}_\mathrm{imp}
  &+ w \sum^{L}_{i=-L}\sum_\sigma i^2\hat{n}_{i\sigma}.
\end{align}
Here, \(J \) is the hopping amplitude, \(\hat{c}^\dagger_{i\sigma}\) (\(\hat{c}_{i\sigma}\)) creates (annihilates) an itinerant fermion with spin \(\sigma \) at site \(i\), 
\(V\) is the spin-exchange interaction between an itinerant fermion and a localized spin at site 0, 
\(\hat{\bm{s}}_i = (\hat{s}^x_i, \hat{s}^y_i, \hat{s}^z_i) \) are spin operators of an itinerant fermion at site \(i\) and 
each component is defined as \(\hat{s}^\gamma = (1/2)\sum_{\alpha \beta} \hat{c}^\dagger_{i \alpha}\sigma^\gamma_{\alpha \beta}\hat{c}_{i\beta}\) with Pauli matrices \(\bm{\sigma}^\gamma \), \(\hat{\bm{S}}_\mathrm{imp} = (\hat{S}^x_\mathrm{imp}, \hat{S}^y_\mathrm{imp}, \hat{S}^z_\mathrm{imp})\) are spin operators of the localized spin, \(w\) is the amplitude of a trap potential, and \(\hat{n}_{i \sigma} = \hat{c}^\dagger_{i\sigma}\hat{c}_{i \sigma}\).
We choose the same parameter with the main text for the performance test, i.e., \(V/J=1.0\), \(L=75\), \(w/J = 0.001\), the number of particles are set to nine, and the total magnetization is set to zero.
We also choose the same unitary operator with the main text for \(\hat{U}\):
\begin{align}
    \hat{U} = {\left[\mathrm{e}^{-\mathrm{i}\hat{H}_\mathrm{even}/J}\mathrm{e}^{-\mathrm{i}\hat{H}_\mathrm{odd}/J}\right]}^4.
\end{align}
Here, \(\hat{H}_\mathrm{odd}\) (\(\hat{H}_\mathrm{even}\)) is the Hamiltonian linking odd (even) bonds of the Kondo model~\eqref{eq:hamiltoinan}.
On-site terms are equally divided to \(\hat{H}_\mathrm{odd}\) and \(\hat{H}_\mathrm{even}\).

The benchmark of the modified METTS algorithm is performed by comparing the time required to reach the stationary value of the total energy with the original METTS algorithm with the fixed total particle number.
An inverse temperature \(\beta \) is set to \(2.5 J^{-1}\).
All calculations are performed on a single thread of Intel Xeon E5--2683 v4 processor and use the same pseudo-random number sequence.

\begin{figure}
    \centering
    \includegraphics[width=0.75\linewidth]{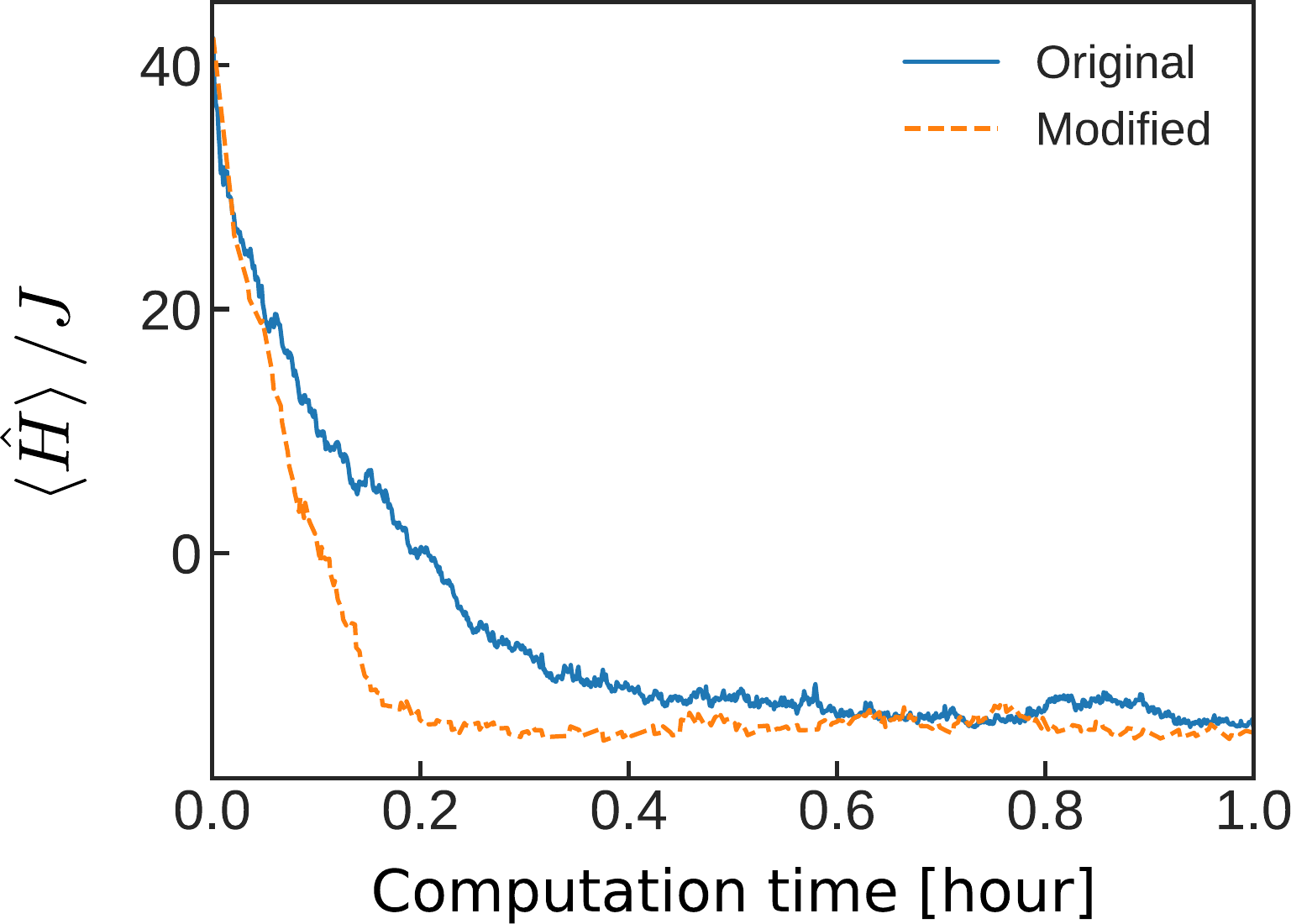}
    \caption{The comparison of computation time required to reach the stationary distribution between the original (blue solid) and modified (orange dashed) METTS algorithms.~\label{fig:energy-time}}
\end{figure}
Figure~\ref{fig:energy-time} represents the chronological sequence of the total energy \(\braket{\hat{H}}\) in the original (blue solid line) and modified (orange dashed line) METTS algorithm.
From this data, one can infer how fast the Markov chain reaches their stationary distribution.
Although it is difficult to point out precisely when the stationary distributions are realized, the modified METTS algorithm approaches the stationary value twice or three times as fast as the original METTS algorithm: The applications of the Trotter gates reduces the computation time indeed.

\begin{figure}
    \centering
    \includegraphics[width=0.75\linewidth]{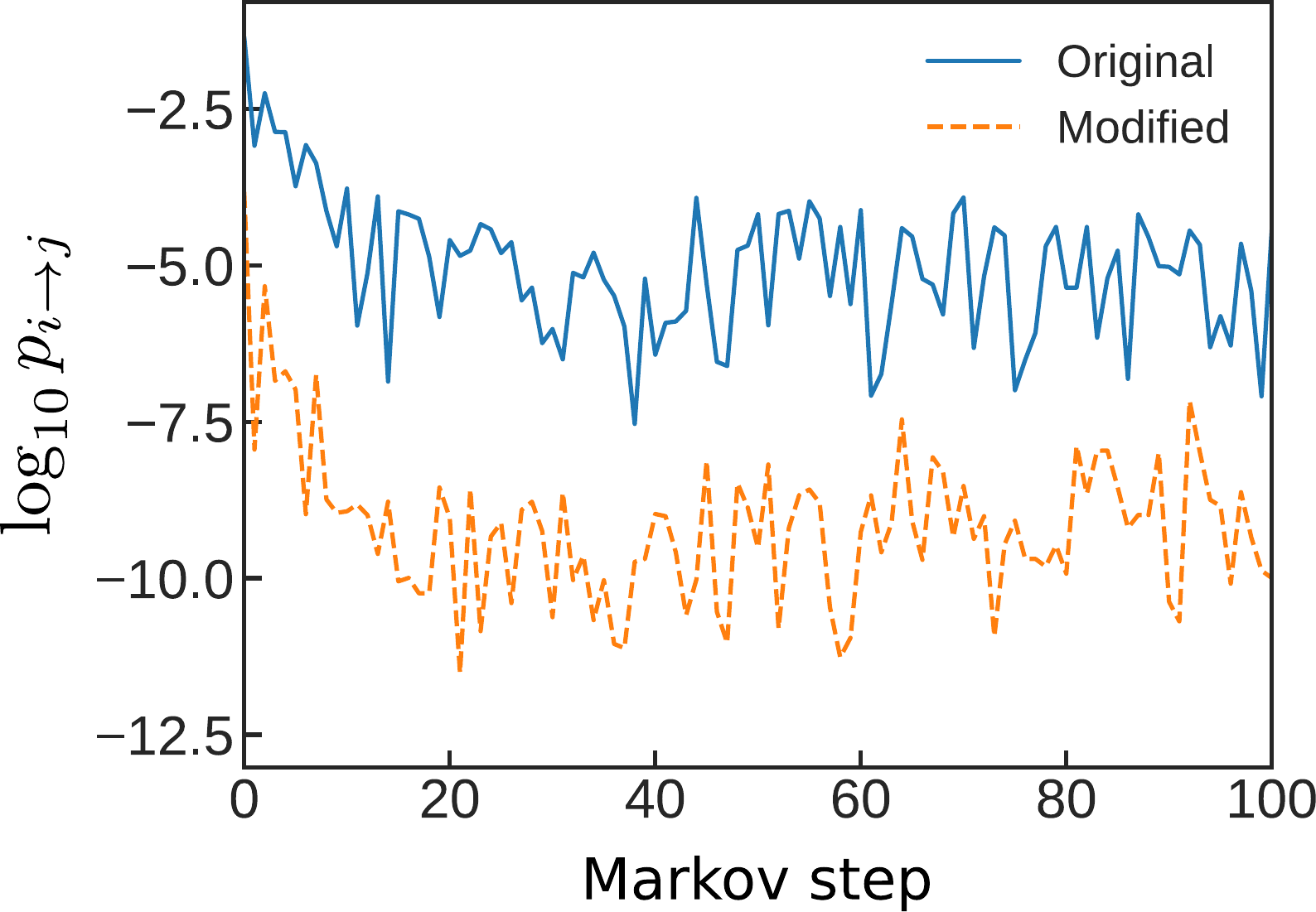}
    \caption{The selected transition probability for first hundred Markov steps in the original (blue solid) and modified (orange dashed) METTS algorithms.~\label{fig:prob-step}}
\end{figure}

The improvement given by the applications of the Trotter gates are also visible in the selected transition probabilities adopted at each Markov step.
Figure~\ref{fig:prob-step} shows the transition probability selected in first hundred Markov steps.
In the original METTS algorithm (blue solid line), the typical order of the transition probability is \(10^{-5}\).
On the other hand, the typical order in the modified METTS algorithm (orange dashed line) is \(10^{-10}\): Roughly speaking, the modified METTS algorithm select a next state from \(10^{5}\) times larger region of the Hilbert space compared to the original METTS algorithm.
Because of the larger region, the autocorrelation of samples in the modified METTS algorithm is much smaller than that of the original METTS algorithm and the small autocorrelation leads to numerically efficient simulations shown in Fig.~\ref{fig:energy-time}.

\section{The Kondo temperature and the Kondo length}
In the main text, we define the Kondo temperature \(T_\mathrm{K}\) as a temperature at which the spin correlation \(\mathrm{Re}\sum_i\braket{\hat{S}^+_\mathrm{imp}\hat{s}^-_i}\) becomes -0.25, namely the half of the maximal singlet value.
In this section, we show that the Kondo temperature defined by the spin correlation value has exponential dependence on the spin-exchange interaction \(V\) likewise the ordinary Kondo temperatures given by the perturbative renormalizetion group analysis for around \(V/J = 1\).
On the other hand, there are noteworthy differences between the perturbative Kondo temperature and the spin-correlation Kondo temperature for small \(V\).
We also discuss these differences from the view point of the Kondo length and the inhomogeneous trap potential.

From the perturbative renormalizetion group analysis, the Kondo temperature for small \(V\rho \) is given as~\cite{hewson_kondo_1997} 
\begin{align}
  \label{eq:perturbative_Tk}
  T_\mathrm{K, p} = \frac{D}{2} \sqrt{V \rho} \exp \left( -\frac{1}{V \rho} \right),
\end{align}
where \(D\) is the band width and \(\rho \) is the density of states at the Fermi level.
With the one-dimensional cosine band \(E(k) = -2 J \cos k a\) and the particle density \(n\), \(D=4J\) and \(\rho = 1/(\pi J \sin \frac{\pi n}{2})\).
Here, \(a\) is the lattice constant.
Since the Kondo temperature indicates the crossover region, its order of magnitude and dependence on the interaction are important rather than its precise value.

\begin{figure}
  \centering
  \includegraphics[width=0.9\linewidth]{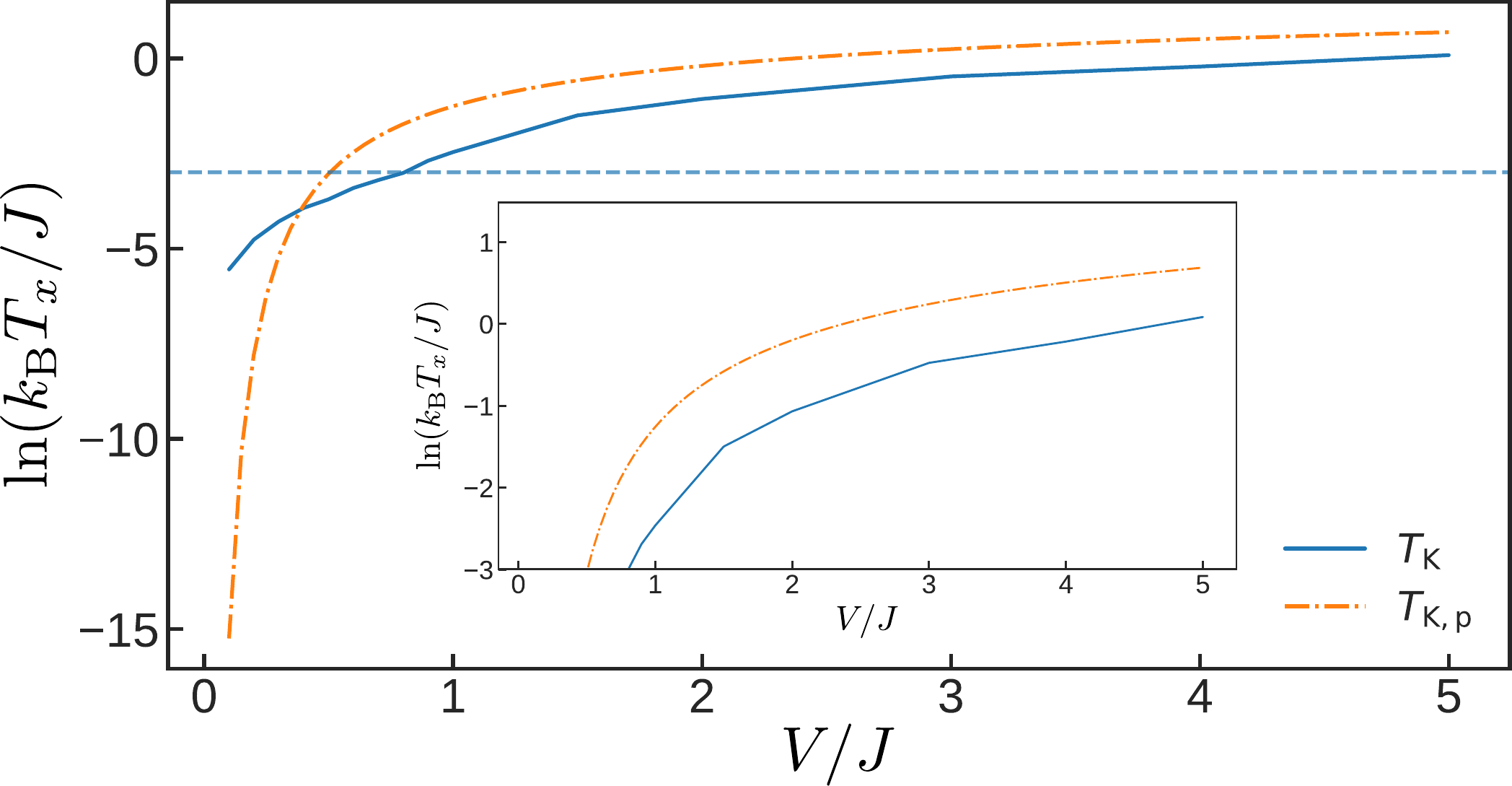}
  \caption{The comparison of the Kondo temperatures determined from the spin correlation \(T_\mathrm{K}\) and obtained from the perturbative approach \(T_\mathrm{K, p}\)~\eqref{eq:perturbative_Tk}. The horizontal dashed line represents the Kondo temperature where the size of the Kondo cloud becomes comparable to the region where itinerant atoms are distributed. Inset: The enlarged figure for the region above the horizontal line.~\label{fig:Tk}}
\end{figure}

In the settings used in the main text, the particle density of the trapped itinerant atoms is around \(0.3\) and the atoms are distributed over 40 sites (See Fig.\@ 1 of the main text).
Figure~\ref{fig:Tk} shows the spin-exchange interaction dependence of the Kondo temperatures determined from the spin correlation and obtained from the perturbative approach~\eqref{eq:perturbative_Tk}.
For around \(V/J = 1\), the two Kondo temperatures agree well except some factor which might come from the arbitrariness of the spin-correlation value used to define the Kondo temperature.
This agreement supports that the ``Kondo temperature'' determined from the spin correlation surely behaves as the ordinary Kondo temperature.

On the contrary, for small \(V/J \lesssim 0.8\), the two Kondo temperatures show a discrepancy and the discrepancy becomes larger as \(V/J\) decreases.
This behavior can be understood from the viewpoint of the Kondo length \(\xi_\mathrm{K}\), the size of the Kondo screening cloud whose scale is given as \(\hbar v_\mathrm{F} / (k_\mathrm{B}T_\mathrm{K})\)~\cite{sorensen_scaling_1996}.
Here, \(v_\mathrm{F}\) is the Fermi velocity which is given as \(v_\mathrm{F} = (2 Ja / \hbar)\sin\frac{\pi n}{2}\) for the one-dimensional cosine band with the particle density \(n\). Then, \(\xi_\mathrm{K} \sim a / (k_\mathrm{B}T_\mathrm{K} / J)\) for \(n \approx 0.3\).
In the settings of the main text, the itinerant atoms are distributed over 40 sites and the localized atoms are located at the center of the distributed region.
Thus, \(\xi_\mathrm{K} \sim 20 a\) is the achievable longest value in this settings.
In terms of the Kondo temperature, \(T_\mathrm{K} \sim 0.05 J / k_\mathrm{B}\) at \(V/J \simeq 0.8\) is the simulatable lowest Kondo temperature and the dashed horizontal line in Fig.~\ref{fig:Tk} represents this threshold value.
Therefore, the large discrepancy in \(T_\mathrm{K} \lesssim 0.05 J/k_\mathrm{B}\), or equivalently \(V/J \lesssim 0.8\) can be understood as a signal of the finite size effects induced by the inhomogeneous trap potential, and the fact that such an interpretation can be provided also supports that the ``Kondo temperature'' \(T_\mathrm{K}\) obtained from the spin correlation can be regarded as the ordinary Kondo temperature as long as the Kondo length is sufficiently small compared to the size of the region where itinerant atoms are distributed.

\begin{figure}
  \centering
  \includegraphics[width=0.9\linewidth]{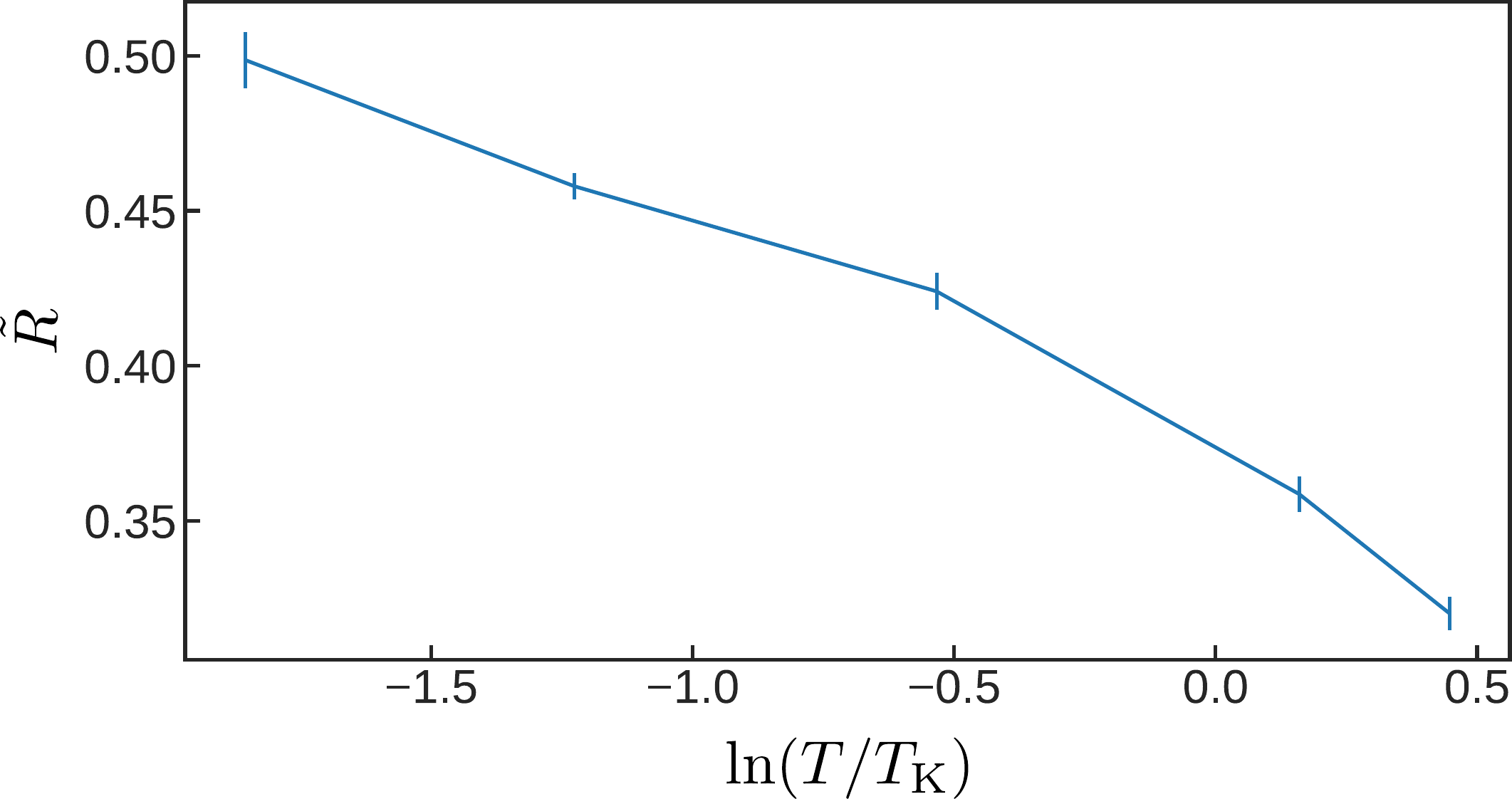}
  \caption{Scaled temperature dependence of \(\tilde{R}\) in the antiferromagnetic Kondo model.~\label{fig:R_Tk}}
\end{figure}

With the Kondo temperature determined from the spin correlation, we plot the scaled temperature dependence of the ``resistance'' \(\tilde{R}\).
Figure~\ref{fig:R_Tk} represents the resistance \(\tilde{R}\) in the antiferromagnetic Kondo model as a function of \(\ln(T/T_\mathrm{K})\).

\section{Relation between resistance and the maximum center-of-mass velocity}
In this section, we relate the resistance \(R\) and the quantity
\begin{align}
  \tilde{R} = 1 - \frac{v_\mathrm{p}}{v_\mathrm{f}}
\end{align}
by regarding the difference of kinetic energy between undamped and damped dipole oscillations as the Joule heat~\cite{danshita_universal_2013}.
Here, \(v_\mathrm{p}\) is the peak velocity of the center-of-mass (COM) motion during the damped dipole oscillation and \(v_\mathrm{f}\) is the maximum COM velocity of the undamped dipole oscillation.

After the quarter of the dipole oscillation period \(t_\mathrm{q}\), the COM velocity reaches maximum value and the difference of the kinetic energy between undamped and damped oscillations at this time point is given by
\begin{align}
  E_\mathrm{diff} = \frac{1}{2}m_\mathrm{tot}(v^2_\mathrm{f} - v^2_\mathrm{p})
  \label{eq:E_diff}
\end{align}
with the total mass \(m_\mathrm{tot}\).
Assuming \(\tilde{R} \ll 1\), or equivalently \((v_\mathrm{f} - v_\mathrm{p})/v_\mathrm{f} \ll 1\), the energy difference~\eqref{eq:E_diff} can be approximated to 
\begin{align}
  E_\mathrm{diff} &= m_\mathrm{tot} v^2_\mathrm{f} \left[1 - \frac{v_\mathrm{p}}{v_\mathrm{f}}-\frac{1}{2}{\left(\frac{v_\mathrm{f}-v_\mathrm{p}}{v_\mathrm{f}}\right)}^2\right] \nonumber \\ 
  &\approx m_\mathrm{tot} v^2_\mathrm{f} \tilde{R}.
  \label{eq:E_diff_approx}
\end{align}

If we regard the effects of interactions on the dipole oscillation as those from a resistor, its resistance \(R\) is obtained by identifying the kinetic energy difference to the Joule heat generated during the quarter period, in short,
\begin{align}
  E_\mathrm{diff} = RI^2 t_\mathrm{q}
  \label{eq:E_diff_Joule}
\end{align}
where \(I\) is the time-averaged particle current over the quarter period which is almost proportional to \(v_\mathrm{f}\), \(I \sim \alpha v_\mathrm{f}\) with some constant \(\alpha \).
From Eqs.~\eqref{eq:E_diff_approx} and~\eqref{eq:E_diff_Joule}, one can derive
\begin{align}
  \tilde{R} \sim \frac{\alpha^2 t_q}{m_\mathrm{tot}}R,
\end{align}
and thus \(\tilde{R}\) is approximately proportional to the resistance \(R\) under the assumption that \(\tilde{R} \ll 1\).